\documentclass[fleqn,usenatbib]{stile/mn2e}

\usepackage{txfonts}
\usepackage[T1]{fontenc}
\usepackage{ae,aecompl}

\usepackage{graphicx}

\usepackage{amsmath}
\usepackage{amsfonts}
\usepackage{amssymb}

\usepackage{xcolor}

\usepackage{breakcites}
\usepackage{hyperref}
\usepackage{footnote}

\usepackage{comment}

\usepackage{xfrac}        

\def\nat{Nature}

\def\mnras{MNRAS}
\def\apj{ApJ}
\def\apjl{ApJL}
\def\apjs{ApJS}
\def\aap{A\&A}

\def\araa{ARA\&A}

\def\pasj{Pub. Astron. Soc. Japan}

\newcommand{\quotes}[1]{``#1''}

\def\gtsima{$\; \buildrel > \over \sim \;$}
\def\ltsima{$\; \buildrel < \over \sim \;$}
\def\gsim{\lower.5ex\hbox{\gtsima}}
\def\lsim{\lower.5ex\hbox{\ltsima}}
\def\gtsima{$\; \buildrel > \over \sim \;$} 
\def\ltsima{$\; \buildrel < \over \sim \;$} \def\gsim{\lower.5ex\hbox{\gtsima}} 
\def\lsim{\lower.5ex\hbox{\ltsima}} 
\def\simgt{\lower.5ex\hbox{\gtsima}} 
\def\simlt{\lower.5ex\hbox{\ltsima}}

\def\be{\begin{equation}}
\def\ee{\end{equation}}

\usepackage{hyperref}	
\hypersetup{colorlinks=true,linkcolor=blue,citecolor=blue,filecolor=blue,urlcolor=blue}
\usepackage{lipsum}
\newcommand{\cii}{[C\,{\sc ii}]}

\newcommand{\oiii}{[O\,{\sc iii}]}

\newcommand{\target}{A1689-zD1}
\usepackage{ragged2e}

\definecolor{referee}{rgb}{0,0,0}

\graphicspath{{plots/}}                 

\author[Tom Bakx et al.]{Tom J. L. C. Bakx$^{1,2}$\thanks{E-mail: bakx@a.phys.nagoya-u.ac.jp (Nagoya University)},
Laura Sommovigo$^{3}$, 
Stefano Carniani$^{3}$,
Andrea Ferrara$^{3}$,\newauthor
Hollis B. Akins$^{4}$,
Seiji Fujimoto$^{5,6}$,
Masato Hagimoto$^{1}$,
Kirsten K. Knudsen$^{7}$,\newauthor
Andrea Pallottini$^{3}$,
Yoichi Tamura$^{1}$ and
Darach Watson$^{5,6}$
\\
$^{1}$ Division of Particle and Astrophysical Science, Graduate School of Science, Nagoya University, Aichi 464-8602, Japan.\\
$^{2}$ National Astronomical Observatory of Japan, 2-21-1, Osawa, Mitaka, Tokyo 181-8588, Japan. \\
$^{3}$ Scuola Normale Superiore, Piazza dei Cavalieri 7, 56126, Pisa, Italy \\
$^{4}$ Department of Physics, Grinnell College, 1116 Eighth Ave., Grinnell, IA 50112, USA \\
$^{5}$ Cosmic Dawn Center (DAWN), Copenhagen, Denmark \\
$^{6}$ Niels Bohr Institute, University of Copenhagen, Jagtvej 128, Copenhagen, Denmark \\
$^{7}$ Department of Earth and Space Sciences, Chalmers University of Technology, Onsala Space Observatory, SE-43992 Onsala, Sweden \\
}
\date{Accepted 2021 September 3. Received 2021 September 3; in original form 2021 August 10.}

\pubyear{2021}

\begin{document}

\newcommand{\LS}[1]{({\bf \color{teal} LS: #1})}
\newcommand{\TB}[1]{({\bf \color{purple} TB: #1})}
\def\S21{\citetalias{Sommovigo2021}}

\title[Band\ 9 continuum in the EoR]{Accurate dust temperature determination in a $z=7.13$ galaxy}

\label{firstpage}
\pagerange{\pageref{firstpage}--\pageref{lastpage}}

\maketitle

\begin{abstract}
    We report ALMA Band\ 9 continuum observations of the normal, dusty star-forming galaxy \target{} at $z = 7.13$, resulting in a $\sim$4.6\,$\sigma$ detection at $702$\,GHz. 
    For the first time these observations probe the far infrared (FIR) spectrum shortward of the emission peak of a galaxy in the Epoch of Reionization (EoR).
    Together with ancillary data from earlier works, we derive the dust temperature, $T_{\rm d}$, and mass, $M_{\rm d}$, of \target{} using both traditional modified blackbody spectral energy density fitting, and a new method that relies only on the \cii{} $158\ \mathrm{\mu m}$ line and underlying continuum data. The two methods give $T_{\rm d} = (42^{+13}_{-7}, 40^{+13}_{-7}$)\,K, and $M_{\rm d} = (1.7^{+1.3}_{-0.7}, 2.0^{+1.8}_{-1.0})\,\times{}\,10^{7} \,M_{\odot}$. Band\ 9 observations improve the accuracy of the dust temperature (mass) estimate by $\sim 50$\% (6 times). The derived temperatures confirm the reported increasing $T_{\rm d}$-redshift trend between $z=0$ and $8$; the dust mass is consistent with a supernova origin. Although \target{} is a \textit{normal} UV-selected galaxy, our results, implying that $\sim$85\% of its star formation rate is obscured, underline the non-negligible effects of dust in EoR galaxies.
\end{abstract}


\section{Introduction}
Atacama Large Millimeter/submillimeter Array (ALMA) observations have revealed the presence of dust in galaxies approaching the epoch of reionization (EoR; e.g., \citealt{Capak2015,Willott2015,Barisic2017,Laporte2017}). This was somewhat surprising, since UV studies mapping out the cosmic star-formation rate density (SFRD) to $z \sim 10$ suggested a dearth of dust at the high-redshift end based on the blue UV slopes of low-stellar mass high-$z$ galaxies ($\beta_{\rm UV}$; e.g., \citealt{Finkelstein2015,Bouwens2016}). Initially, the strong FIR emission at $z>7$ revealed by ALMA observations was attributed to the presence of unexpectedly large dust masses ($M_{\rm d}$) in the observed high-$z$ galaxies, which was hard to reconcile with known dust production mechanisms that operate on that timescale (predominantly SN and grain growth; see \citealt{Lesniewska2019} and references therein for the latest constraints).

This resulted in the so-called {\it dust budget crisis}, which also impacted star-formation history (SFH) estimates of high-redshift galaxies \citep[e.g.,][]{Mawatari2020,RobertsBorsani2020}. The stringent constraints on SNe dust production, coupled with the large deduced dust masses at $z> 7$, required very early stellar populations originating at $z \sim 14$ \citep{Tamura2019}.
However, the conclusions on the dust masses were heavily dependent on the assumed (cold) dust temperatures ($T_{\rm d} \sim 30-40\ \mathrm{K}$) for these high-$z$ sources, since in most cases only a single data point was available in the FIR continuum. Recent observations (e.g., \citealt{Bakx2020}) and theoretical studies (e.g. \citealt{Behrens2018,Sommovigo2020}) have suggested the presence of warm dust in several high-$z$ galaxies ($T_{\rm d} >60\ \mathrm{K}$), alleviating the large dust mass requirements set by their observed $L_{\rm FIR}$ ($M_{\rm d} \propto T_{\rm d}^{\rm -(4+\beta_d)}$ at fixed $L_{\rm FIR}$, where typically $1.0<\beta_{\rm d}<3.0$). Unfortunately, the large uncertainties on derived $T_{\rm d}$ at high-$z$ still hinder accurate SFH studies.

Partially due to the lack of knowledge on the dust temperature at high-$z$, the total fraction of obscured star-formation beyond $z > 4$ is also largely unknown \citep{Novak2017,Casey2018,Bouwens2020,Gruppioni2020,Schouws2021,Talia2020,Zavala2021}. 
This has strong implications for the cosmic SFRD; for example, some of these recent works suggest that {\color{referee} there is no steep drop-off in SFRD at $z>3$} (e.g., \citealt{Gruppioni2020}), which could indicate that we might be underestimating the contribution of highly obscured systems to the SFRD at $z>3$ due to the bias towards UV bright objects. On top of that, most studies calculate the obscured star-formation rates and far-infrared luminosities of single sources either by assuming a dust temperature, and/or by scaling directly from the infrared excess (${\rm IRX} = L_{\rm FIR}/L_{\rm UV}$)-$\beta_{\rm UV}$ relation. Both approaches suffer from the inherent uncertainty in dust temperature (since obscured SFR and IRX both scale with $T^{4+\beta_{\rm d}}$).

Moreover, the validity of IRX-$\beta_{\rm UV}$ relation at high-$z$ demands that the UV and dust-emitting regions to be co-spatial, relying on the absorbed UV emission to be re-emitted at far-infrared wavelengths. However, observations suggest the possibility of spatial separation between these regions in several sources {\color{referee} at $z=4-6$ (e.g., \citealt{Faisst2017})} and at $z \sim 7-8$ sources (e.g., \citealt{Carniani2017,Laporte2019} and Tamura et al. in prep.). In fact, this {\it spatial separation} scenario between UV and IR is also supported by theoretical studies and simulations \citep{Behrens2018,2019Cochrane,2019Liang,Sommovigo2020}. A deviating IRX-$\beta_{\rm UV}$ relation would impact the results of galaxies at high-$z$ \citep{LeFevre2020, Fudamoto2020}, and will impact re-emission studies (e.g.,  MAGPHYS; \citealt{daCunha2008} and CIGALE; \citealt{Boquien2019}) which will be prevalent in the ALMA\,+\,JWST era.

In this letter, we use the band\ 9 observations to estimate the dust properties of a $z = 7.1$ galaxy from the spectrum directly in order to measure the obscured star-formation directly. We describe the source and data in Section~\ref{sec:Ch2}, the fitting techniques in Section~\ref{sec:Ch3} and the implications in Section~\ref{sec:Ch4}.\footnote{Throughout this paper, we assume a flat $\Lambda$-CDM cosmology with the best-fit parameters derived from the \textit{Planck} results \citep{Planck2015}, which are $\Omega_\mathrm{m} = 0.307$, $\Omega_\mathrm{\Lambda} = 0.693$ and $h = 0.678$.}

\begin{table}
	\caption{Continuum and fitting properties of \target{}}
	\label{tab:data}
	\begin{tabular}{llll} 
\hline	
		& $\lambda$\,[mm] & F$_{\nu}^{\rm int}$\,[$\mu$Jy]$^{\dagger}$ & Reference \\
\hline
Band 9 & 0.427 & 154 $\pm$ 37 & This work \\
Band 8 & 0.728 & 180 $\pm$ 39 & \citet{Inoue2020}  \\
Band 7 & 0.873 & 143 $\pm$ 15 & \citet{Knudsen2017}   \\
Band 6 & 1.33 & 60 $\pm$ 11  & \citet{Watson2015}    \\
		\hline
		& SED fit & \cii{}-based & SED fit (no B9) \\ 
		\hline \vspace{0.1cm}
$T_{\rm d}$ [K]     & 42$^{+13}_{-7}$           & $40_{-7}^{+13}$           & $38_{-13}^{+35}$ \\ \vspace{0.1cm}
$\beta_{\rm d}$     & 1.61$^{+0.60}_{-0.75}$    & {\it 2.03}\,$^{\ddagger}$   & 1.78$^{+0.55}_{-0.97}$        \\ \vspace{0.1cm}
$M_{\rm d}$ [$10^{7}\ \mathrm{M_{\odot}}$]$^{\dagger}$ & 1.7$^{+1.3}_{-0.7}$ & $2.0^{+1.8}_{-1.0}$ & $2.4^{+11}_{-1.9}$ \\ \vspace{0.1cm}  
$L_{\rm FIR}$ [$10^{11}\ \mathrm{L_{\odot}}$]$^{\dagger}$ & 1.9$^{+0.5}_{-0.4}$ & 2.2$^{+4.1}_{-1.0}$ & 1.5$^{+3.0}_{-0.6}$ \\ \vspace{0.1cm}
${\rm log_{10}\ IRX}$ & 1.0$^{+0.1}_{-0.1}$ & 1.0$^{+0.5}_{-0.3}$  & 0.8$^{+0.5}_{-0.2}$ \\ 
$M_{\rm d}$/SN [$\mathrm{M_{\odot}}$]$^{\dagger}$ & 0.4$^{+0.3}_{-0.1}$ &  $0.6^{+0.6}_{-0.3}$ & 0.8$^{+3.0}_{-0.7}$  \\
\hline
	\end{tabular}
	\raggedright \justify \vspace{-0.2cm}
\textbf{Notes:} $\,^{\dagger}$ Corrected for the magnification assuming $\mu = 9.3$ from \cite{Knudsen2017}. $^{\ddagger}$ $\beta_{\rm d}$ is fixed to 2.03. 

\end{table}

\section{Target and Observations}
\label{sec:Ch2}
A1689-zD1 was identified in \cite{Bradley2008} as a bright (m$_{\rm AB} \sim 25$) $z>7$ galaxy. Due to the foreground galaxy cluster (A1689; \citealt{Struble1999}), it is magnified\footnote{While $\mu$ is high, there is only little shear, and we do not account for any differential lensing effects in this paper} by $\mu \simeq 9.3$ \citep{Knudsen2017}. Its intrinsic UV magnitude indicates it is a sub-$L$* galaxy representing the bulk of galaxies at $z = 7$ \citep{Ono2018}. Band\ 6 observations at 1.3\,mm by \cite{Watson2015} reported the first detection of dust beyond redshift 7, and indicated an intrinsic star-formation rate of $\sim 12$\,$\rm M_{\odot}$\,$\rm yr^{-1}$. Notably, the estimated dust mass of this {\it normal} galaxy (assuming 35\,K) was found in tension to star-formation history and dust production estimates in \cite{Lesniewska2019}.

In this letter, we combine the existing data on \target{} reported in \cite{Watson2015}, \cite{Knudsen2017} and \cite{Inoue2020} with 
archival band\ 9 data from (Program ID: 2019.1.01778.S, P.I. D. Watson), see Table~\ref{tab:data}. We use the \cii{} luminosity as reported in Knudsen et al. in prep., which is (6.1 $\pm$ 0.7)$\ \times\ 10^8$\,L$_{\odot}$, and use their value for spectroscopic redshift, $z_{\rm spec} = 7.13$.

\begin{figure}
	\includegraphics[width=\columnwidth]{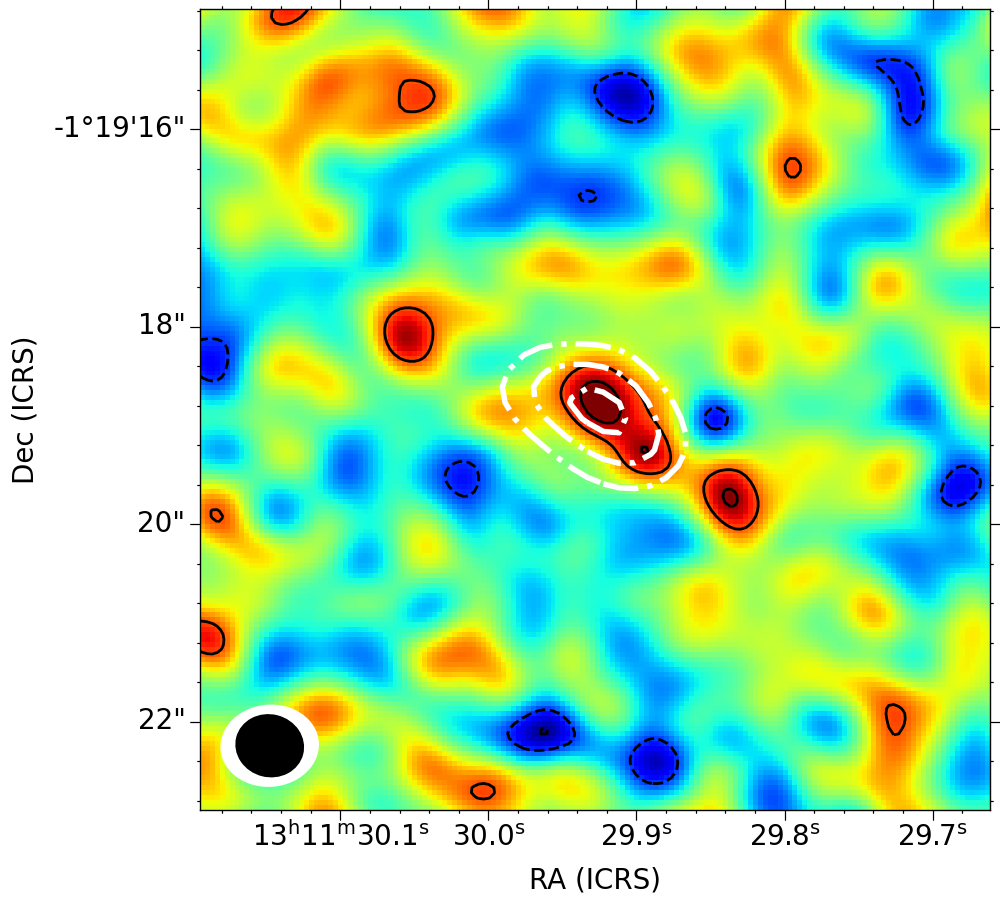}
    \caption{The tapered band\ 9 data ({\it background and black contours}; drawn at -3, -2, 2 and 3\,$\sigma$) is shown against the band\,8 continuum emission ({\it white contours}; drawn at 5, 7 and 10\,$\sigma$). The continuum emissions appears co-spatial, and we find a 4.6$\sigma$ dust detection in band\,9. }
    \label{fig:continuumImage}
\end{figure}

For the band\ 9 \citep{Baryshev2015} data, the source was observed for 95\,min. in baselines ranging from 14 to 312\,m. The lower and upper sidebands covered the contiguous frequency ranges of 690.4 to 697.6 and of 706.5 to 713.6\,GHz. We assume a typical flux accuracy of 10\%. The continuum image is produced with \texttt{CASA} pipeline version 5.6.1-8 \citep{McMullin2007}, using natural weighting, a taper of 0.5 arcseconds, and excluding any channels within 1000\,km/s of the \oiii{} 52$\mu$m emission at 711.4\,GHz. Figure~\ref{fig:continuumImage} shows the resulting image with a 0.61 by 0.67 arcsecond beam with a beam position angle of 75 degrees, with an r.m.s. level of 210\,$\mu$Jy\,beam$^{-1}$. 
Using \texttt{CASA}'s \texttt{IMFIT} routine, we spatially integrate the emission using a 2D Gaussian profile. This results in a flux of 1.43 $\pm$ 0.31\,mJy ($\sim 4.6 \sigma$; excluding calibration flux), with an apparent (or lensed) beam-deconvolved size of 0.81 $\pm$ 0.26 by 0.38 $\pm$ 0.22 arcsec at a position angle of 44 $\pm$ 38 degrees. {\color{referee} The emission appears co-spatial to the UV-emission seen in \cite{Knudsen2017}, although we leave further discussion of this to Knudsen et al. in prep.}

\section{Methods}
\label{sec:Ch3}

\subsection{Spectral fitting}
\begin{figure}
	\includegraphics[width=\columnwidth]{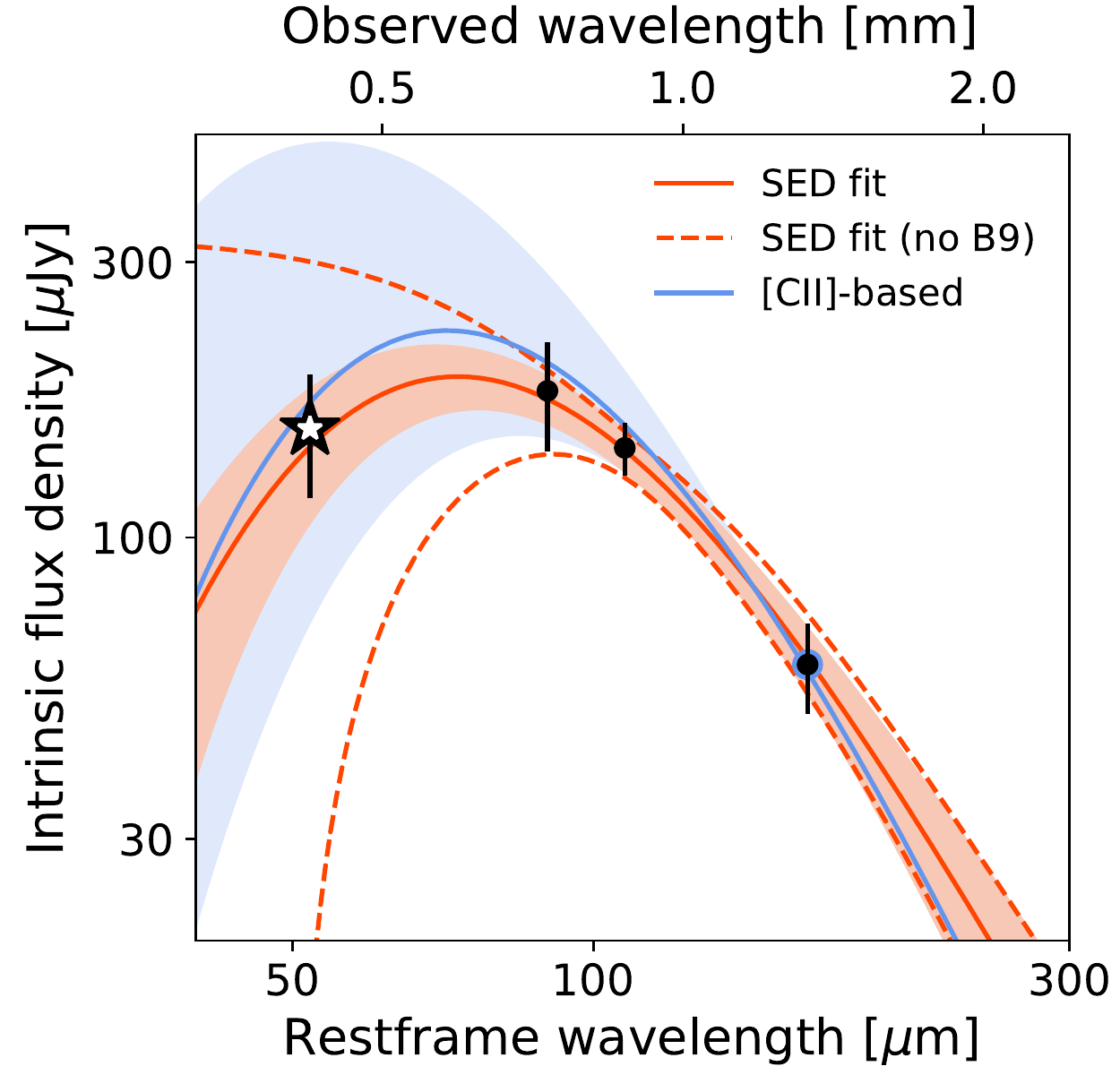}
    \caption{We fit a modified black-body (\textit{red line and fill}) to the observed data points of A1689-zD1, including the band\ 9 data point ({\it star}). The \cii{}-based spectrum (\textit{blue line and fill}) is fit solely to the 158\,$\mu$m continuum data point ({\it blue}), and it predicts a consistent galaxy spectrum, providing confidence in the \cii{}-based method for this specific source even at 50\,$\mu$m rest-frame. For comparison, the {\it dashed red lines} show the spread in SEDs fitted without band\ 9 data, which results in a twice larger error in dust temperature, {\color{referee} and a sixfold increase in the error in dust mass.}
    }
    \label{fig:spectrum}
\end{figure}

Figure\ \ref{fig:spectrum} shows the modified black-body {\color{referee} (eq. 8 in \citealt{Sommovigo2021})} fitted to the continuum points reported in Table~\ref{tab:data}. We use equations~12 and 18 from \cite{daCunha2013} to account for the heating of dust by and decreasing contrast against the CMB, respectively. We approximate the dust mass absorption coefficient ($\kappa_{\rm \nu}$) as $\kappa_{\rm \star} \left( \nu / \nu_{\star}\right)^{\beta_{\rm d}}$, with ($\kappa_{\rm \star}, \nu_{\star})$ as (10.41\,cm$^2$/g, 1900\,GHz) from \cite{Draine03}.
We use the \texttt{emcee} MCMC-fitting routine, and allow $M_{\rm d}$, $T_{\rm d}$ and $\beta_{\rm d}$ to vary freely using flat priors, resulting in a {\color{referee} magnification-corrected} dust mass of 1.7$^{+1.3}_{-0.7}$\,$\times{} 10^{7}$\,M$_{\odot}$, a dust temperature of 42$^{+13}_{-7}$\,K and a $\beta_{\rm d}$ of 1.61$^{+0.60}_{-0.75}$. We note that the spectrum appears well-represented by a single modified black-body. For comparison, we also fit the spectral energy distribution (SED) without band\ 9 data, with an upper limit on $\beta_{\rm d}$ of 2.5 to ensure convergence, and find significantly larger errors across the board.
{\color{referee} If we take a fiducial $\beta_{\rm d} = 1.8$ (e.g., \citealt{Casey2012,Faisst2020}), we find a more accurate dust temperature of 39$^{+4}_{-5}$\,K, however there is no improvement on the error of dust mass (2.0$^{+1.4}_{-0.7} \times 10^7$\,M$_{\odot}$) nor luminosity (1.7$^{+0.5}_{-0.4} \times 10^{11}$\,L$_{\odot}$). The accuracy of these later parameters thus depends solely on observational uncertainties, indicating that we fully trace the dust emission in this source.}

\subsection{Dust temperature from \cii{} emission}
We use the novel method proposed in \cite{Sommovigo2021} to derive the dust temperature in galaxies, based on the combination of $1900\ \mathrm{GHz}$ continuum and the overlying \cii{} line emission. We provide a brief summary of this method below; for further details and verification of this method on 19 local galaxies, three galaxies at $z \simgt 4$, and a $z \sim 6.7$ simulated galaxy, we refer to \citet{Sommovigo2021}.

We relate the observed \cii{} luminosity to the total dust mass via the gas mass and a gas-to-dust ratio (assumed to scale linearly with the metallicity, which is justified down to $Z \simlt 0.1\ Z_{\odot}$, see \citealt{james2002, 2007ApJ...657..810D,galliano2008,Leroy_2011}). The gas mass and \cii{} luminosity are related through a conversion factor $M_{\rm gas}= \alpha_{\rm CII}\ L_{\rm CII}$. This conversion factor $\alpha_{\rm CII}$ is analytically derived from the combination of the de Looze relation \citep{delooze14} and the Kennicutt–Schmidt relation \citep[][hereafter, KS]{Kennicutt1998}. Two parameters are added in the expression for $\alpha_{\rm CII}$ in order to account for both (i) the expected offset from the KS-relation (i.e., the burstiness of the SF of a galaxy parametrized by $\kappa_{\rm s}$) and (ii) the observed larger extension of \cii{} with respect to stellar emission at high-$z$ \citep[up to 1.5-3 times larger;][]{carniani:2017oiii,carniani2018clumps,Carniani20,matthee2017alma,Matthee_2019,2019ApJ...887..107F,2020arXiv200300013F,Fujimoto2021,2020A&A...633A..90G,HerreraCamus2021}. 

We fit a modified blackbody to derive the dust temperature using the neighbouring continuum emission at $\sim$1900\,GHz rest-frame wavelength, {\color{referee} assuming a fixed $\beta_{\rm d} = 2.03$, which is based on the \cite{Draine03} predictions for the Milky Way and the Small Magellanic Cloud}. Within this fitting routine, both the burstiness parameter ($\kappa_{\rm s}$) and the metallicity are largely uncertain. In order to constrain the dust temperature, two broad physical constraints are placed: (i) The dust mass cannot exceed the maximal dust mass producible by supernovae (SNe), assuming all the SNe metal yield ($\sim2\ \mathrm{M_{\odot}}$ per SN) ends up locked into dust grains; (ii) The dust-obscured star formation\footnote{This IR luminosity-to-SFR conversion factor, $1.73 \times 10^{-10}\ \mathrm{M_{\odot} yr^{-1}/L_{\odot}}$, is valid for a Salpeter $1-100\ \mathrm{M_{\odot}}$ IMF, which we assume consistently throughout the paper} 
\citep{Kennicutt1998,Madau2014}, cannot significantly (by 1 order of magnitude) exceed the SFR deduced from \cii{} using the relation from \cite{delooze14} for starbursts. 
Applying our method to A1689-zD1, we find a dust temperature $T_{\rm d}=40_{-7}^{+13}\ \mathrm{K}$ and mass of $M_{\rm d}=2.0^{+1.8}_{-1.0}\ \times 10^{7}\ \mathrm{M_{\odot}}$. These values are obtained assuming a wide range of values for the metallicity $Z=0.2-1\ \mathrm{Z_{\odot}}$, and the burstiness parameter $\kappa_{\rm s}=1-50$ \citep[][]{2021arXiv210605279V}. {\color{referee} For our further discussion of dust production mechanisms, we note that the removal of the dust production constraint does not influence the derived quantities.}

\section{Implications} 

\label{sec:Ch4}

\begin{figure}
	\includegraphics[width=1.0\linewidth]{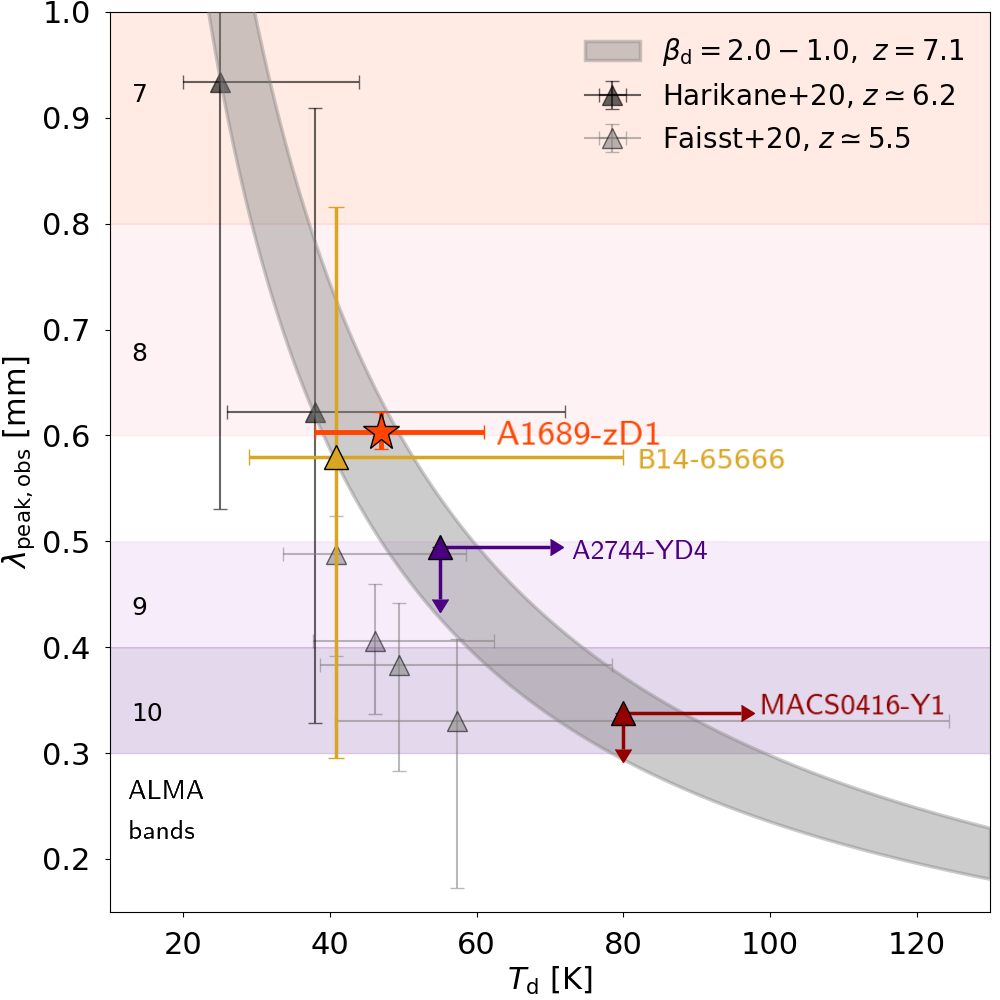}
    \caption{Observed peak wavelength $\lambda_{\rm peak,obs}$ is shown against dust temperature $T_{\rm d}$ for a given dust emissivity index, $\beta_{\rm d}$. The grey shaded region shows $\lambda_{\rm peak,obs}(T_{\rm d})$ at redshift $z=7.1$ for $\beta_{\rm d}=2.0-1.0$. We show the sources with reported dust temperatures beyond $z>5$ (\citealt{Faisst2020, Harikane2019,Sugahara2021,Bakx2020,Laporte2019}). The shaded regions show the wavelength ranges probed by ALMA bands $7-10$. Without band\ 9, {\color{referee} we cannot probe the FIR peak on both sides} and accurately estimate $T_{\rm d}$ through SED fitting, while for lower-redshift observations band\ 10 might even be required to accurately trace the SED.}
    \label{fig:obspeak}
\end{figure}
The dust temperature and mass estimates from the \cii{}-based method agree with the results from the direct SED fitting, which adds confidence to the method from \cite{Sommovigo2021}. 
As shown in Table~\ref{tab:data}, band\ 9 observations reduce the uncertainty in the dust temperature by $\sim 50\%$, which translates to much-improved estimate on the far-infrared luminosity and dust mass estimate. 
In Figure~\ref{fig:obspeak}, we show the observed peak emission wavelength ($\lambda_{\rm peak,obs}$) of galaxies at $z > 5$ against our current best-estimates for $T_{\rm d}$ and $\beta_{\rm d}$. To guide the eye, we include the trend of $\lambda_{\rm peak,obs}$ with $T_{\rm d}$ for $\beta_{\rm d} = 1-2$ at $z = 7.1$. We also overlay the wavelengths of the ALMA bands\ 7 through 10. We calculate this $\lambda_{\rm peak,obs}$ using
\begin{equation}
\begin{split}
\frac{\lambda_{\rm peak,obs}}{\mathrm{mm}} & =  
\frac{ 14.42\ (1+z)\ (T_{\rm d}/K)^{-1}} {W(-a\ e^{-a})+a}\  ,\nonumber
\end{split}
\end{equation}

\noindent
where $a=3+\beta_{\rm d}$ and $W$ is the Lambert $W$ function. This is the wavelength at which the continuum spectrum $F_{\rm \nu}$ peaks in frequency units (e.g., Figure\ \ref{fig:spectrum}). This is an important distinction to keep in mind when visualising $\lambda_{\rm obs, peak}$ from the analogy to Wien's law, which provides the peak of the spectrum when reported in wavelength units $F_{\lambda}$.

Particularly for galaxies at lower redshifts and at higher temperatures, short-wavelength observations are crucial to estimate the dust temperature, 
whereas band\ 8 might be able to probe the emission peak for cold $z > 8$ galaxies. 
In the foreseeable future, the high bands of ALMA (9 and 10) are the only instrument capable of probing this regime, 
until such missions as the Origins Space Telescope\footnote{https://asd.gsfc.nasa.gov/firs/docs/} \citep{Meixner2019}.

\begin{figure}
	\includegraphics[width=1.0\linewidth]{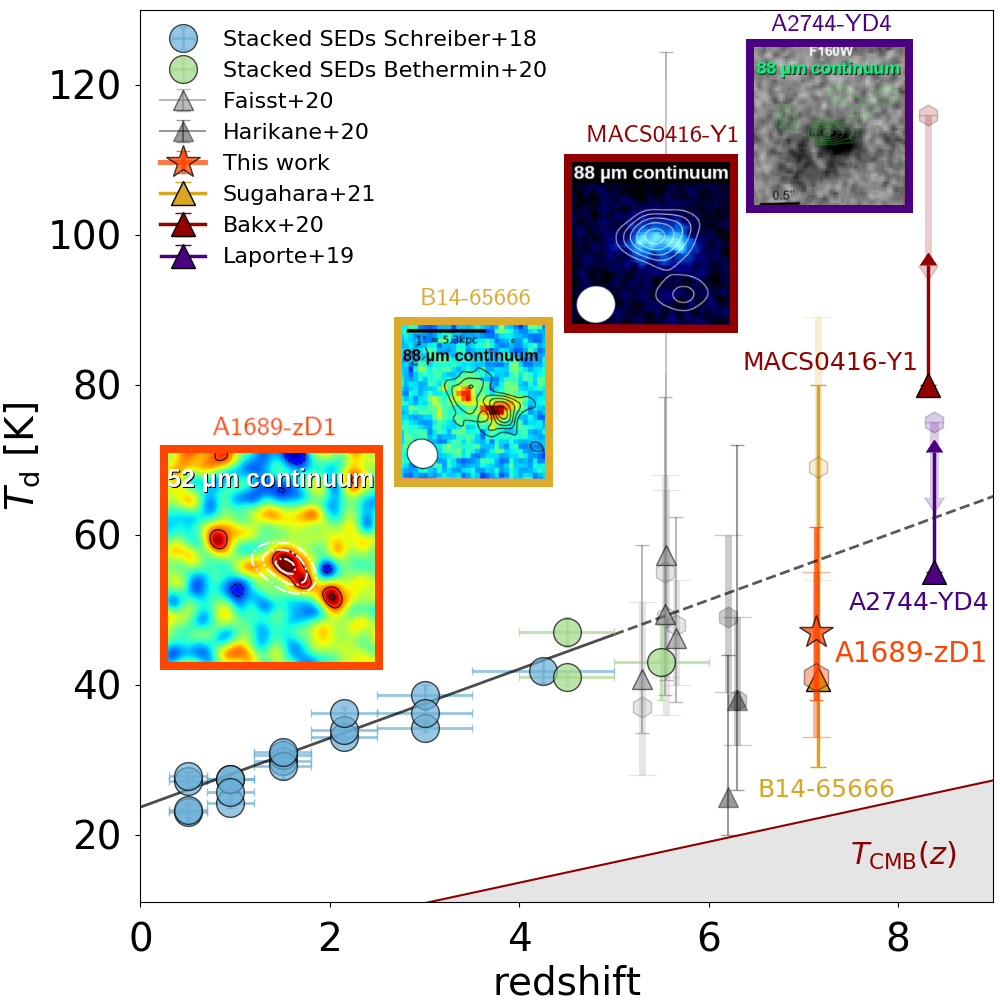}
    \caption{Dust temperature $T_{\rm d}$ in \quotes{normal} (main-sequence) galaxies as a function of redshift. The newest $T_{\rm d}$ estimates for \target{} are shown in {\it red} ({\it star} for SED-fit and {\it hexagon} for the \cii{}-based result). Dust temperatures obtained from stacked SEDs ({\it blue and green circles}) increase linearly with redshift up to $z = 6$. We highlight all the continuum detected sources at $z>7$ with small post-stamps, and include their estimated $T_{\rm d}$ based on both SED fits ({\it triangles}) and the \cii{}-based method ({\it hexagons}; \citealt{Faisst2020, Harikane2019,Sugahara2021,Bakx2020,Laporte2019}). The addition of band\ 9 data significantly reduces the uncertainty on the dust temperature of this source with respect to the other high-$z$ sources, which are not observed in that band.}
    \label{Tdz}
\end{figure}
Accurate estimates of the the dust-obscured fraction of the star-formation rate require strong constraints on the dust temperature, as $\mathrm{SFR}_{\rm obsc} \propto L_{\rm FIR} \propto M_{\rm d} T_{\rm d}^{4+\beta_{\rm d}}$.
Our band\ 9 observations confirm that this relatively-cold ($T_{\rm d}$\,$\sim 40 - 60$\,K) system has a very large obscured fraction of the SFR around $\sim$85\% ($\mathrm{SFR}_{\rm obs}=33 \pm 9\ \mathrm{M_{\odot} yr^{-1}}$, whereas\footnote{derived using the magnification-corrected $L_{\rm UV}/10^{10}\ \mathrm{L_{\odot}}=2.28 \pm 0.1$ \citep{Hashimoto2019}, and the UV luminosity-to-unobscured SFR conversion factor in \cite{Madau2014}.} $\mathrm{SFR}_{\rm UV}=5.7 \pm 0.3\ \mathrm{M_{\odot} yr^{-1}}$), even though it was selected to be UV-bright. {\color{referee} The dust-obscured ratio is higher than the 61\% found for the typically more-massive ALPINE survey \citep{LeFevre2020,Faisst2020,Fudamoto2020,Bethermin2021,Khusanova2021} at $z=5.5$, although this 61\% is expected to decrease with higher redshift. Albeit extreme, our dust-obscured ratio is in line with recent results (both theoretical and observational) suggesting that we might have been underestimating the dust-obscured contribution to the total SFR in $z>4$ galaxies \citep[see e.g.][]{Novak2017,Gruppioni2020}. On the other hand, some studies of similarly-massive, UV-bright sources at very high-$z$ ($z \sim 7$; e.g., \citealt{Bouwens2021Rebels, Schouws2021}) have so far failed to detect dust continuum at 158\,$\mu$m in half of their sources, even though their average stellar mass is similar to those of \target{}. These undetected sources might have low dust contents, but that does not guarantee low obscured star-formation fractions, since it is possible that this dust is warm and is mainly emitting at wavelengths shorter than the currently observed ones (mainly $158\ \mathrm{\mu m}$ rest-frame). In fact, while the continuum around $158\ \mathrm{\mu m}$ of MACS0416\_Y1 \citep{Tamura2019,Bakx2020} has yet to be seen, its spectrum is indicative of a similar obscured fraction to \target{} (i.e., 94 to 85\% for $\beta_{\rm d} = 1.5$ to 2), even though Y1's UV-observed stellar mass is one order of magnitude lower than \target{}.

The selection towards UV-bright sources might also bias towards lower fractions of obscured-to-total star-formation rate. With the discovery of so-called optically-dark galaxies (e.g., \citealt{Simpson2014,Franco2018,Hatsukade2018,Wang2019,Yamaguchi2019,Williams2019,Algera2020b,Romano2020,Talia2020,Toba2020,Umehata2020,Zhou2020,Shibuya2021,Smail2021}), we know of the existence of galaxies without detections in optical wavelengths at high redshift. These sources, by definition, have exceedingly high obscured fractions and might well account for a substantial fraction of the SFRD at high redshift \citep{AlcaldePampliega2019,Gruppioni2020,Zavala2021}. The typical obscured star-formation rate fraction across all $z > 7$ galaxies might thus be higher than predicted by UV-selected samples alone, with for example \cite{Gruppioni2020} predicting an increase in SFRD of 17\% at $z = 5$ by this population.
}

Recently, attempts to quantify dust obscuration at high-$z$ have used a linear scaling between dust temperature (and $L_{\rm IR}$ given a fixed $\beta_{\rm d}$) and redshift \citep[see e.g.][]{Schreiber18,Bouwens2020,Vijayan2021}. Other recent works have suggested that this linearly increasing $T_{\rm d}-z$ trend flattens at $z>4$ \citep{2019Liang,Faisst2020}. In Figure~\ref{Tdz} we show the reported linear evolution of the dust temperature with redshift, adding our latest results for \target{}, and where available, include the results from the method in \cite{Sommovigo2021}. 
The observed dust temperature for \target{} is compatible with both a flattening \citep{2019Liang,Faisst2020} and a linear \citep{Schreiber18} $T_{\rm d}-z$ evolution. Meanwhile, the exceedingly-large scatter in $T_{\rm d}$ at the highest redshifts (particularly at $z>7$) prevents us from reaching a definitive conclusion on this observed evolution. Much of this scatter is due to observational limitations, and only through further short-wavelength observations of galaxies beyond $z > 7$ can we distinguish the possible scenarios. Part of the scatter could also be due to a larger source-to-source variation in $T_{\rm d}$, {\color{referee} which is for example seen by the large diversity of galaxies among the typically more-massive ALPINE sources \citep{LeFevre2020}. Such source-to-source variation can only be identified by larger unbiased samples looking at the dust-obscured star-formation at high redshift. Here we note that an increased intrinsic scatter in dust temperature would significantly boost the resulting dust-obscured star-formation rate, given their strong dependence of star-formation rate on dust temperature, similar to an Eddington-type bias.}

Due to the large obscured faction of the SFR in \target{}, one might naively expect that this galaxy also contains an exceedingly large dust mass. Instead, the dust mass derived from SED fitting implies a dust yield of $y_{\rm d} = 0.4^{+0.3}_{-0.1}\ \mathrm{M_{\odot}}$ per SN. This estimate is almost an order of magnitude more accurate than the one derived without band\ 9 data, and most importantly, it is consistent with latest SN dust production constraints by \cite{Lesniewska2019} based on the expected number of SNe given its stellar mass estimate. They find at most a $y_{\rm d} = 1.1\ \mathrm{M_{\odot}}$ per SN, derived in the extreme case of no dust destruction/ejection. We note that SN yield is still highly debated, with other works suggesting that dust destruction processes might only spare $0.1\ \mathrm{M_{\odot}}$ per SN \citep[e.g.,][]{Matsuura2015,Matsuura2019,Slavin2020}. {\color{referee} In this extreme case, inter-stellar medium grain growth \citep[][]{Mancini2015,Michalowski2015} or more exotic dust production mechanisms might well be required at $z > 7$, such as dust produced in supershells \citep[e.g.,][]{MartinezGonzalez2021} or in the wake of Wolf-Rayet stars \citep[e.g.,][]{Lau2021}.}

\bibliographystyle{mnras}
\bibliography{bibMake}

\section*{Data Availability}
The data underlying this article will be shared on reasonable request to the corresponding author.

\section*{Acknowledgements}
We would like to thank the anonymous referee for their insightful comments and suggested additions.
This paper makes use of the following ALMA data: ADS/JAO.ALMA 
2011.1.00319.S, 
2012.1.00216.S, 
2013.1.01064.S, 
2016.1.00954.S, and
2019.1.01778.S.
TB and YT acknowledge funding from NAOJ ALMA Scientific Research Grant Numbers 2018-09B and JSPS KAKENHI No. 17H06130. 
LS, SC, AF, AP acknowledge support from the ERC
Advanced Grant INTERSTELLAR H2020/740120 (PI: Ferrara).
The Cosmic Dawn Center is funded by the Danish National Research Foundation under grant No. 140.
This project has received funding from the European Union’s Horizon 2020 research and innovation program under the Marie Sklodowska-Curie grant agreement No. 847523 ‘INTERACTIONS’.
DW and SF are supported by Independent Research Fund Denmark grant DFF--7014-00017
Any dissemination of results must indicate that it reflects only the
author’s view and that the Commission is not responsible for any use
that may be made of the information it contains.

\label{lastpage}
\end{document}